\documentstyle[aps,prd,twocolumn,tighten,psfig]{revtex}
\newcommand{\beq}{\begin{equation}}
\newcommand{\eeq}{\end{equation}}
\newcommand{\bea}{\begin{eqnarray}}
\newcommand{\eea}{\end{eqnarray}}

\newcommand{\ie}{{\em i.e.\/}}

\newcommand{\et}{{\em et al.\/}}

\newcommand{\wh}{wormhole\/}
\newcommand{\btr}{\bigtriangledown}
\begin{document}
\draft
\title{ Electromagnetic waves in a wormhole geometry}
\author{S. E. Perez Bergliaffa
\thanks{Electronic mail: \tt santiago@lafex.cbpf.br} and K. E. Hibberd}
\address{Centro Brasileiro de Pesquisas F\'{\i}sicas \\
Rua Dr.\ Xavier Sigaud 150, 22290-180 Urca, Rio de Janeiro, RJ -- Brazil} 
\date{\today}
\renewcommand{\thefootnote}{\fnsymbol{footnote}}
\twocolumn[
\hsize\textwidth\columnwidth\hsize\csname@twocolumnfalse\endcsname 
\maketitle

\begin{abstract}
\hfill{\small\bf Abstract}\hfill\smallskip
\par
We investigate the propagation of electromagnetic waves through a static wormhole. It is shown that the problem can be reduced to a one-dimensional Schr\"odinger-like equation with a barrier-type potential. Using numerical methods, we calculate the transmission coefficient as a function of the energy. We also discuss the polarization of the outgoing radiation due to this gravitational scattering.
\end{abstract}
\smallskip\mbox{}]
\footnotetext[1]{Electronic mail: \tt santiago@lafex.cbpf.br}
\renewcommand{\thefootnote}{\arabic{footnote}}

\section{Introduction}

Since the pioneering article by Morris and Thorne \cite{motho}, Lorentzian wormholes have attracted a lot of interest in the literature.  
Let us recall that a wormhole is a solution to Einstein's equations that can be understood as a ``handle'' connecting either two universes or two distant places in the same universe.  A number of static and dynamic wormholes have been found both in General Relativity and in alternative theories of gravitation 
(see \cite{hochvis1,hochvis2} and references therein).  It has also been shown that under particular circumstances this geometry permits the formation of closed timelike curves (CTC's) \cite{timemach}. 

One of the most striking features of this field configuration is that it needs matter with negative energy density as a source \cite{motho}.  Hochberg and Visser, and Ida and Hayward have analyzed in detail the issue of the violation of the energy conditions both in static \wh s \cite{hochvis1} and in a completely general 
(\ie\/  non-symmetric and time dependent) traversable \wh\/ \cite{hochvis2,ida}.  Their analysis shows that the null energy condition (NEC) must be violated in both cases.  Since it is believed that all classical matter acting as a source of gravitation satisfies the NEC, only quantum phenomena can, 
in principle, be responsible for NEC violations.  In fact, it has been known for some time that there are states in Quantum Field Theory that may violate the energy conditions \cite{epstein}.  Many examples of systems that can violate these conditions due to quantum effects can be found in \cite{visserlib};
one such example is the conformally coupled scalar field. Indeed, Hochberg \et\/ \cite{hochberg} found the first self-consistent \wh\/ solution in semiclassical General Relativity with quantized conformally coupled scalar matter as a source
\cite{bar}. Another instance in which quantum effects are essential has been presented recently by Kim and Lee \cite{kimlee}. Using a two-dimensional dilaton-gravity at the one-loop level, they showed that a wormhole can be the final state of an evaporating black hole. This result agrees with previous considerations of Hayward \cite{sean}, and might be considered as an example of a process to bring a wormhole into existence. Recently, Krasnikov \cite{kras} showed that there exist static wormholes with the vacuum stress-energy tensor of the neutrino, the electromagnetic or the massless sclar field as a source of Einstein's equations.

In spite of these results, the issue of the existence of large amounts of NEC violating matter has not yet been resolved.  However, one can take a complementary view by assuming that wormholes exist and work out possible consequences.  This line of reasoning was started by Frolov and Novikov \cite{frono}, who studied nontrivial effects arising in a \wh\/ that interacts with an external electromagnetic or gravitational field.  In a subsequent paper \cite{frono2}, they used wormholes
 as tools for studying the interior of black holes.  Later, Gonzalez-Diaz studied some astrophysical consequences of the existence of these objects \cite{gondiaz}.  More recently, Cramer \et \cite{cramer} considered the unusual features of the lensing of light caused by an object with negative mass (which could be interpreted as one of the mouths of a \wh ).  As a natural application of this idea, Torres \et\/ \cite{torres} considered the microlensing caused by wormholes on light coming from  distant active galactic nuclei.  They showed that these events resemble certain types of Gamma Ray Bursts (GRB's) and set an upper bound on the negative mass density existing in the universe in the form of wormholes. The GRB's produced by microlensing by wormholes present a definite feature that differentiate them from those associated with fireballs \cite{fireball}. Namely, they always appear in pairs called FRED-antiFRED.
A subsequent article by Anchordoqui {\em et al} \cite{doqui2} investigated profiles of GRB's from the BATSE database, searching for observable signatures of natural wormholes.  They could identify at least one event that may be associated with this mechanism.

From an observational point of view, it is also important to study the propagation of different types of perturbations in the geometry associated with these objects.  This study was initiated by Kar and Sahdev \cite{kar1}, and Kar \et\/ \cite{kar2}, who studied the reflection and transmission of massless scalar waves in the presence of an ultrastatic wormhole.  In this article we consider the propagation of electromagnetic waves in the same geometry.  We will study some properties of the outgoing radiation modified by the gravitational field of the wormhole.

We should mention at 
this point an article that is at first sight related to ours. In \cite{clement},
Cl\'ement discussed the problem of the scattering of scalar and electromagnetic waves in an Ellis geometry (which is in fact a wormhole). Following 
the ideas of Wheeler's geometrodynamics \cite{wheeler}, his aim was to show that wormholes are particle-like objects. Consequently, the wormhole structure in \cite{clement} was to be important only at a microscopic level, while in this work we are considering macroscopic wormholes.

The structure of the paper is as follows.  In the next section we transform Maxwell's equations to a form that is convenient to study the propagation of electromagnetic waves in curved spacetimes.  Following this, we introduce the metric of a static wormhole in which we will study the electromagnetic perturbations.  Section IV shows how the problem is equivalent to a one-dimensional problem in the presence of a potential barrier.  We also present numerical results obtained for the transmission coefficient.  Section V investigates the polarization of the outgoing radiation.  We conclude with a summary of our results.

\section{Maxwell's equations in a gravitational field}

We begin with the equations that govern the propagation of electromagnetic waves in a gravitational background \cite{notation}
\bea
F^{\mu\nu}_{\;\; ;\nu} = 4\pi J^\mu, ~~~~~~~
F_{\mu\nu ;\sigma} +F_{\nu\sigma ;\mu}+
F_{\sigma\mu ;\nu} = 0. 
\label{max1}
\eea
where $J^\mu$ is the current four-vector.
We shall see below that these equations can be rewritten in a more convenient way \cite{volkov}. 
In a given coordinate frame such that \\ 
$ds^2 = g_{\mu\nu}dx^\mu dx^\nu$, define
$$H^{\mu\nu} \equiv \sqrt{-g}\;F^{\mu\nu}, ~~
\mbox{ and } ~~ I^\mu \equiv \sqrt{-g}\; J^\mu .$$ 
In terms of these tensors, Maxwell's equations are given by
\bea
H^{\mu\nu}_{\;\;\; ,\nu}=4\pi I^\mu, ~~~~~~ F_{\mu\nu ,\sigma} +
F_{\nu\sigma ,\mu}+ F_{\sigma\mu ,\nu} = 0. \nonumber 
\eea
Note that these equations are similar to Maxwell's equations in flat spacetime.
The background geometry has not disappeared, but manifests itself in the constituitive equations
$H^{\mu\nu} = \sqrt{-g}\; g^{\mu\rho}g^{\nu\sigma}F_{\rho\sigma}$.
To be specific, in a Cartesian coordinate 
system the following decomposition of the tensors is possible:
\beq
F_{\mu\nu}\rightarrow (\vec E, \vec B), ~~~~ 
H^{\mu\nu}\rightarrow (-\vec D, \vec H), ~~~~
J^\mu \rightarrow(\rho , \vec J\;). 
\label{carte}
\eeq
In this notation, Maxwell's equations explicitly take the form they have in Euclidean spacetime:
\begin{eqnarray}
\vec\bigtriangledown .\vec B & =& 0, ~~~~~~ \vec\bigtriangledown\wedge\vec E = -\frac{\partial\vec B}{\partial t}, \nonumber \\
\vec \bigtriangledown . \vec D &=& 0 ,~~~~~~
\vec\bigtriangledown\wedge\vec H = \frac{\partial \vec D}{\partial t} + 4\pi \vec j. 
\end{eqnarray}
With the decomposition given in Eq.(\ref{carte}) the constituitive relations can be written as
\beq
D_i = \epsilon_{ik}E_k - (\vec G\times\vec H)_i \nonumber
\eeq
and
\beq
B_i = \mu_{ik}H_k+(\vec G\times \vec E)_i,\nonumber
\eeq
where 
\beq
\epsilon_{ik}=\mu_{ik} = -\sqrt{-g}\;\frac{g^{ik}}{g_{00}}
\label{perm}
\eeq
and
\beq
G_i = -\frac{g_{0i}}{g_{00}}. \nonumber
\eeq
We can now see that Maxwell's equations in a gravitational background are formally equivalent 
to the equations of an electromagnetic field in a flat spacetime in the presence of a medium. 
Generally, this medium is bi-anisotropic.  In the following, we shall restrict our considerations to a medium 
characterized by diagonal dielectric and magnetic permeabilities :
\beq
\epsilon_{ik} = \mu_{ik} \equiv n\;\delta_{ik}.
\label{indice}
\eeq
In this case, the refraction index $n$ corresponds to a static spacetime \cite{foot}.
Defining the vectors
$$
\vec F^{\pm} \equiv \vec E \pm i\vec H , ~~~~~ \vec S^\pm \equiv \vec D \pm i\vec B,
$$
it is easily shown that Maxwell's equations can be recast as
\beq
\vec\btr\wedge\vec F^{\pm} = \pm i \frac{\partial \vec S^\pm}{\partial t}, 
~~~~~~~
\vec\btr . \vec S^\pm = 0.
\label{maxeqs}
\eeq
Using the fact that $\vec D = \epsilon \vec E$, and $\vec B = \mu \vec H$, the first of these equations reduces to
\beq
\vec\btr\wedge \vec F^\pm = \pm i\;n~ \frac{\partial\vec F^\pm}{\partial t}.
\label{mainmax}
\eeq
In the next section we describe the background geometry for the wormhole in which the electromagnetic perturbations will be studied. 

\section{The geometry}

As the background metric, we shall adopt that of a static wormhole, given by 
\cite{motho}
\bea
ds^2 =-e^{2\phi(r)}dt^2 + \frac{dr^2}{\left(1-\frac{b(r)}{r}\right)} + r^2 d\Omega ^2, \label{metric}
\eea
where $d\Omega ^2$ is the surface element of $S^2$, $\phi (r)$ is the so-called redshift function 
and $b(r)$ is the shape function.
Here we shall restrict ourselves to the case $b(r) = b_0^2/r$, where $b_0$ is the radius of the throat of the 
wormhole. In order to use the decomposition given by Eq.(\ref{carte}), 
we need to recast the metric in Cartesian coordinates.
First we make the transformation to isotropic coordinates by introducing $r =  f(\rho)$, where
\bea
f(\rho) = \frac{4\rho^2 + b_0^2}{4\rho}. \nonumber
\eea
In these new coordinates, the metric takes the form
\beq
ds^2 = -e^{2\phi(\rho)} dt^2 + A^2(\rho) (d\rho^2 + \rho^2\;d\Omega^2),
\label{metrica2}
\eeq
with
\bea
A(\rho) = \frac{4\rho^2+b_0^2}{4\rho^2}.\nonumber
\eea
The spatial part of the metric can now be written in Cartesian coordinates
by means of the usual definitions $x^1 = \rho \sin\theta\cos\varphi$, $x^2 = \rho\sin\theta\sin\varphi$, $x^3 = \rho\cos\varphi$. 
With this substitution, the metric becomes
\bea
ds^2 =  -e^{2\phi(\rho)} dt^2 + A^2(\rho) (\delta_{ij}dx^i dx^j).
\label{met2}
\eea
From Eqs.(\ref{perm}) and (\ref{indice}) it is easy to show that the refraction index is given by
\bea
n(\rho)= \frac{A(\rho)}{e^{\phi(\rho)}}.\nonumber
\eea
It can be seen from the properties of the metric Eq.(\ref{metric}) that far from the wormhole $n(\rho)$ tends to 1, in which case we will recover Maxwell's equations in free space.

\section{The equivalent one-dimensional problem}

We demonstrate below how the problem of travelling electromagnetic waves in the geometry given by Eq.(\ref{met2}) can be reduced to a one-dimensional Schr\"odinger's 
equation for a particle with unit mass in a given potential.
The following calculations are restricted to the so-called ``ultrastatic case'', 
in which $\phi(\rho) \equiv 0$.  

We begin by developing the Hertz vector $\vec F^\pm$ in series of generalized spherical harmonics \cite{newton}:  
\bea
\vec F^\pm (\vec \rho , t)= \sum_{J,M} \vec F^\pm_{JM} (\vec \rho , t),\nonumber
\eea 
with 
\bea 
\vec F^\pm_{JM} (\vec\rho , t) =
\sum_{\lambda = e,m,o}F^\pm_{JM}(\rho , \omega)\;\vec Y^{(\lambda )}_{JM}(\hat\rho)\;e^{-i\omega t} .\nonumber
\eea 
The superindices $e$ and $m$ refer to transverse modes, while the superindex $o$
refers to longitudinal modes.  Using the
properties of the $\vec Y^{(\lambda )}_{JM}(\hat\rho)$ \cite{newton}, Eq.(\ref{maxeqs}) can be rewritten as 
\beq
-\frac{d}{d\rho}\left(\rho F^{\pm (m)}_{JM} \right) = \pm n\;\omega \rho ~F^{\pm (e)}_{JM} ,
\label{m1}
\eeq 
\beq
\frac{d}{d\rho}\left(\rho F^{\pm (e)}_{JM}\right)-\sqrt{J(J+1)} ~F^{\pm (o)}_{JM} = \pm n\;\omega\rho ~ F^{\pm (m)}_{JM}, 
\label{m2}
\eeq 
and 
\beq -\frac {1}{\rho} \sqrt{J(J+1)}\;F^{\pm (m)}_{JM} = \pm n\;\omega F^{\pm (o)}.  
\label{m3}
\eeq 
At this point, it is convenient to transform $\rho$ to a new variable $x$ \cite{mash1} defined by 
\bea
\frac{dx}{d\rho} = n(\rho). \nonumber 
\eea 
It can be easily seen from the metric given in Eq.(\ref{metrica2}) that the coordinate $x$ is the proper distance,
which is given in terms of $r$ by the relation
\bea 
x = \pm\sqrt{r^2-b_0^2}.\nonumber
\eea 
It is also useful to define the functions 
\bea 
\chi^{\pm(\lambda )}_{JM}(x,\omega) = \rho (x) F^{\pm (\lambda)}_{JM} (\rho(x),\omega). \nonumber
\eea 
Substituting Eqs.(\ref{m1}) and (\ref{m3}) into Eq.(\ref{m2})
and making the transformation $z = x/b_0$
we obtain
\bea 
\frac{d^2\chi^{\pm (m)}_{JM}}{dz^2} + \left[ k^2 -
2U_J(z)\right]\chi^{\pm (m)}_{JM} = 0 ,\nonumber
\eea 
with $k= \omega b_0$.  
The potential $U_J(z)$, is given by 
\beq
U_J(z) = 2J(J+1)\left[ \frac{z+\sqrt{1+z^2}}{(z+\sqrt{1+z^2})^2+1}\right]^2 .
\eeq 
Note that the potential is asymmetric and tends asymptotically to
$0$ as $z\rightarrow \infty$.  This is illustrated in the following figure.
\begin{figure}[h] 
\centerline{\psfig{file=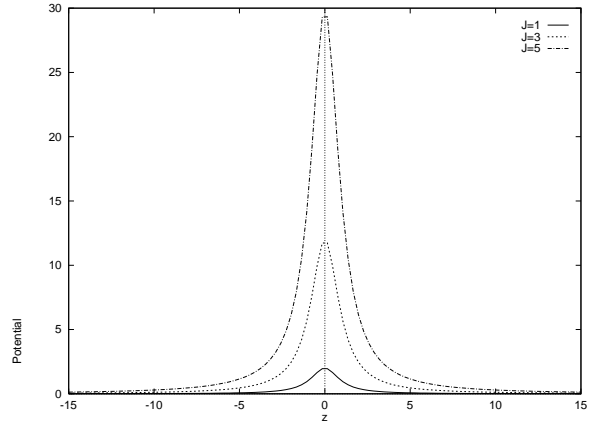,width=8cm,angle=-90}}
\caption{Plot of the potential.} 
\label{potent}
\end{figure}
The potential barrier vanishes for $J=0$, and grows like $J^2$ for large $J$. 
Notice that  $U_J(0)=J(J+1)/2$ is a good estimate for the maximum of the barrier. 
Due to the intricate dependence of the potential on $z$, it was not possible to find an exact analytical expression for the transmission coefficient, $T_J$.
Instead, numerical methods 
based on the work presented in the appendix of \cite{kar2} were employed.  
Our results for the transmission coefficient versus $k$, for the values $J=1,3 \mbox{ and } 5$ are displayed in the following figure.
\begin{figure}[h] 
\centerline{\psfig{file=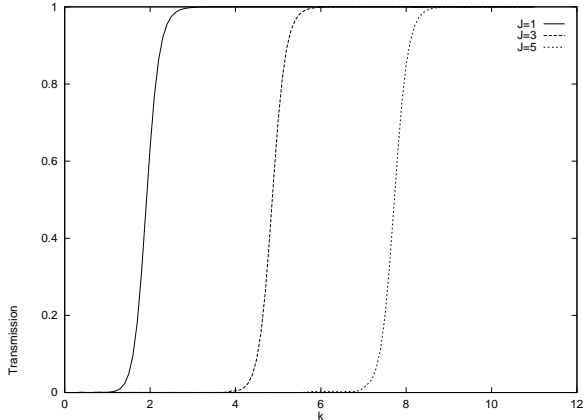,width=8cm,angle=-90}}
\caption{Plot of the transmission coefficient $T_J$ against $k=\omega b_0$.} 
\label{trans}
\end{figure}
$T_J$ gives the amount of radiation that goes through the wormhole. The lower part of the plot shows a well-known feature of this type of scattering: the heigher the potential barrier, the slower $T_J$ grows. However, there is a rapid increment of the transmission coefficient within a small interval of $k$ until it reaches its maximum for every value of $J$.
 
\section{Polarization of the outgoing waves}

Let us now discuss the polarization state of the outgoing waves
that passed through the wormhole's throat. 
From the equations of motion Eq.(\ref{maxeqs}) it can be seen that the left-circularly-polarized radiation ($\vec F^+ = 0$) is decoupled from the right-circularly-polarized photons, so the helicity is conserved.  This remains true even in the case of a stationary spacetime, in which $\vec G \neq 0$. Thus the polarization state of circularly-polarized photons is not altered by the scattering in 
gravitational field of a wormhole \cite{mash2}. Note that this result is valid for any static spacetime (see Eq.(\ref{perm})) and consequently, helicity is conserved for any static wormhole and not just for the one described by Eq.(\ref{met2}).

The case of linearly polarized waves was studied by Mashhoon for a Schwarzschild
black hole \cite{mash1}. His calculations can be easily adapted to the present 
case, but are rather lengthy. Instead, we follow the approach
developed in \cite{nouri}.  The metric of any stationary spacetime can be written as
\beq
ds^2 = h(dt^2 - G_i dx^i)^2-dl^2,
\eeq
with 
\beq
dl^2 = \left(-g_{ij}+\frac{g_{0i}g_{0j}}{g_{00}}\right)
dx^i dx^j.
\nonumber
\eeq
It was shown in \cite{nouri} that the rotation in the polarization plane of light along the path between the source and the observer in the above metric (\ie\/ the so-called Faraday effect) is given by
\beq
\Omega =-\frac{1}{2} \int_{\rm sou.}^{\rm obs.} \sqrt{h}\;\vec B_g . \vec {dl},
\label{cambio}
\eeq
where $\vec B_g$ is the ``gravitomagnetic'' vector given by 
\beq
\vec B_g = \vec\btr \wedge \vec G.
\eeq
In the case of a static wormhole, $\vec G\equiv 0$ and consequently there is no change in the linearly polarized light.

\section{Conclusions}

We have studied here several aspects of the transit of electromagnetic waves through an ultrastatic wormhole. We have shown that the problem can be rewritten in such a way that the curved background geometry is replaced by a medium whose properties are summarized by the refraction index $n(\rho)$. We also showed that the magnetic modes of the electromagnetic field obey a one-dimensional Schr\"odinger-like equation, with an asymmetric barrier-type potential. The transmission coefficient was calculated numerically, and it exhibits some features common to barrier-like potentials.  Namely, in a small interval of $k$, the curve rapidly increases until there is no reflection
and for higher values of $J$, initially the curve rises more slowly. 

We have also demonstrated that the interaction of the radiation with the gravitational field of the wormhole cannot change the polarization state of the radiation. Moreover, this was shown for any static wormhole and not only with the particular case described by metric Eq.(\ref{met2}).

It would be interesting to search for special features of the transmission coefficient in more general types of wormholes. In particular, the existence of bound states (\ie standing waves) for scalar waves was shown in \cite{kar2}. If  resonances with the same characteristics also exist in the case of electromagnetic waves, their position in what could be considered as 
the emission spectrum of the wormhole may give us information on the size of the throat and its shape \cite{kar2}.

One can expect a greater richness of observational consequences if the background wormhole geometry is rotating \cite{theo}. For instance, differential gravitational scattering of polarized radiation can be caused by the  helicity-rotation coupling \cite{mash3}. Finally, to complete the study of perturbations in these spacetimes, it would be necessary to analyze spinor and gravitational perturbations. We intend to address these problems in future works.

\section{Acknowledgments}

We thank S. Joffily, S. Kar, D. Monteoliva, M. Novello, J. Salim, and F. Zyserman for helpful discussions.  S.E. Perez Bergliaffa acknowledges financial support from Conicet (Argentina) and K. Hibberd is supported by CNPq (Conselho Nacional de Desenvolvimento Cient\'{\i}fico e Tecnol\'ogico).


\begin{thebibliography}{99}
\bibitem{foot} In the more general case of stationary spacetimes,
$\vec G\neq 0$, and the constituitive relations will be more complicated.
\end{thebibliography}
\end{document}